# Super resolution and spectral properties for 1D multilayer systems

**A.Mandatori, M.Bertolotti**

*INFM at Dipartimento di Energetica – Università "La Sapienza" di Roma – Via Scarpa 16, 00161, Roma (Italy),+39 06 49941 6541, fax +39 06 442 40 18 3 mario.bertolotti@uniroma1.it*

**Abstract**: we investigate the spectral properties of one-dimensional multilayer structures for the two polarizations TE and TM. We give a physical explanation for the large spatial transmission band that can be obtained with this kind of system, and the correlated super resolution effect. We also suggest a designing approach to build 1D metal-dielectric multilayer structures that have super resolution.

## 1. INTRODUCTION

As first noted by Pendry [1], the amplification of evanescent waves by an ideal negative index material (NIM) layer can lead to subwavelength resolution below the Rayleigh limit. This theoretical prediction was later confirmed by experiments carried out using a flat lens composed of a single silver layer 40-50nm thick [2]. The theory generally requires that the thickness of the metal film should be small compared to the wavelength of the input signal. In fact, unlike an ideal, lossless NIM, which can amplify evanescent waves, silver can only slow down the rate of decay of the evanescent components: in terms of energy flow, the propagation is always characterized by losses. The source is thus generally placed very near to the flat metal lens, typically within $\lambda_0/5$-$\lambda_0/10$, where $\lambda_0$ is the incident wavelength. At such small distances, sub-wavelength resolution can also be obtained with near-field scanning optical microscopy [3-4]. More recently, a new scheme was proposed to increase the distance where super resolution can still be observed [5].

Pendry et al. [6-8] have considered a multilayer structure that improves super-resolution in the near-field zone relative to the single metal layer. The structure consisted of eight periods of alternating metal and air layers that were assumed to have equal thickness (5nm each). The incident wavelength was chosen in a way that the real parts of the permittivity would nearly balance each other, namely $\varepsilon=1$ for air, and $\varepsilon\sim-1$ for silver. Excluding losses, these conditions are approximately satisfied for ultraviolet light in the range between 345 and 360nm, slightly below the plasma frequency of silver ($\lambda_p$~320nm). Although the layered structure displayed improved resolving capabilities, the overall transmittance of propagating modes was nearly identical to the transmittance from a single metal layer 40nm thick, as originally proposed in reference [1]. One obvious drawback in the new system was the much reduced metal layer thickness, down to about 5nm or so, which from a technological point of view is still quite a challenge to achieve.

Another scheme that achieves super-resolution in metal-dielectric structures was recently proposed [9,10] based on thick alternating layers arranged to produce a resonant structure, in a regime where effective medium theory is not applicable. The structure in this case may not transmit evanescent waves and the main underlying physical mechanisms for sub-wavelength focusing are resonance tunnelling, field localization and propagation effects [10]. One important property is its spatial transmission spectrum.

In this paper we perform a study of multilayered structures that are able to exhibit super resolution: we demonstrate that this property is well related to the cavity properties of the structure and that it can involve every kind of 1D multilayer for every input field polarization. By properly choosing the geometry and the refractive index of the layers super resolution can be achieved.

In the next section we present a general discussion of the spatial transmission of unidimentional multilayers and its connection to super-resolution. Section 3 considers the possibility to improve resolution and the final section 4 presents some design criteria to optimize the structure for super-resolution. Conclusions close the paper.

## 2. SPECTRAL PROPERTIES

Before starting to investigate super resolution in 1D multilayer structures and the physical explanation of this phenomenon, we want to discuss some properties of the Fourier transform and of the transmission function of a linear system made by a one-dimensional photonic band gap (PBG) sample.

Between the input $V_i(k_x)$ and the output field $V_o(k_x)$ in the transformation domain there is the relation $V_o(k_x) = T_{PBG}(k_x) V_i(k_x)$, where $T_{PBG}(k_x)$ is the transmission function of the linear system. If we want that the output and the input field are equal, it is necessary that the modulus and the phase of the function $T_{PBG}(k_x)$ be constant for the entire spectral range of the input signal $V_i(k_x)$.

It is clear however that it is impossible to have $T_{PBG}(k_x)$ constant for all real $k_x$ values; there is indeed just the point $k_x / k_0 = 1$ ($k_0 = 2\pi/\lambda$) corresponding to a 90 degrees input where for every linear system we have obligatory $T_{PBG} = 0$. All the values $|k_x / k_0| > 1$ correspond to evanescent waves. Every real system has a transmission that in modulus is strongly different from zero only for a finite range of $k_x$.

The better case of a transmission function that is able to reproduce at the output the input field is represented in fig.1, where m is the maximum spatial frequency that the structure is able to transmit. If now we ignore the two points $k_x / k_0 = \pm 1$ and the two peaks at $\pm\alpha$ which will be discussed later, we can say that the function that better reproduces the input field without changes is a rectangle large 2m and centred in the origin of the $k_x$ axis.

If the input signal has a spectrum completely inside the rectangle, then the output reproduces exactly the shape of the field. If $V_i(k_x)$ has an extension larger than the rectangle, the output signal looses the finesse of the input field, at the high $k_x$ frequencies which are cut by the transmission function of the linear system. This case is exactly what happens in the reality; the transmission function of any PBG cuts the high spatial frequencies. It is obvious that for fixed input signal the larger is the rectangle $|T_{PBG}(k_x)|$ the better will be the reproduction of the input signal. In the case of a PBG it is necessary to find geometries that have transmission function the closest to the one of fig.1 as possible and with the base of the rectangle the largest that is possible.

Now we want to discuss a special transmission function that often approximates the transmission function of many 1D super resolving PBGs: it is the one represented in fig.1 in which two peaks at some values of $k_x / k_0$, say $\pm\alpha$ are shown. Let us first assume that these peaks are two Dirac-$\delta$ functions of area $T_\delta$ centred in the symmetric points $k_x / k_0 = +\alpha$ and $k_x / k_0 = -\alpha$ respectively, with $\alpha > 1$.

The transmission function, disregarding the zero values at $k_x / k_o = \pm 1$, can be described analytically as

$$T_{PBG}(k_x) = \mathrm{rect}_{2mk_0}(k_x) + T_\delta \delta(k_x - \alpha k_0) + T_\delta \delta(k_x + \alpha k_0), \tag{1}$$

where $\mathrm{rect}_T(k_x)$ is the rectangle function with base T centred in the origin. The output field is

$$\begin{aligned} V_o(k_x) &= T_{PBG}(k_x) V_i(k_x) = \\ &= \mathrm{rect}_{2mk_0}(k_x) V_i(k_x) + V_i(\alpha k_0) T_\delta \delta(k_x - \alpha k_0) + V_i(-\alpha k_0) T_\delta \delta(k_x + \alpha k_0) \end{aligned} \tag{2}$$

Writing $V_i(\alpha k_0) = |V| e^{i\Phi}$ and making the Fourier antitransform of eq.(2) we obtain

$$V_o(x) = \Im^{-1}\left\{\mathrm{rect}_{2mk_0}(k_x) V_i(k_x)\right\} + |V| e^{i\Phi} T_\delta e^{i\alpha k_0 x} + |V| e^{-i\Phi} T_\delta e^{-i\alpha k_0 x}, \tag{3}$$

that in a more compact way can be written as

$$V_o(x) = \Im^{-1}\left\{\mathrm{rect}_{2mk_0}(k_x) V_i(k_x)\right\} + 2|V| T_\delta \cos(\Phi + \alpha k_0 x). \tag{4}$$

Eq.(4) describes the electromagnetic field on the output surface of the PBG. It is the sum of two terms, the first one is the Fourier antitransform of the input field after cutting the high spatial frequencies ($|k_x / k_0| > m$), the second one represents an added disturb in the form of a spatial cosine at high frequency (evanescent zone because $\alpha > 1$). The field of eq.(4) must propagate in the free space outside the PBG, and the evanescent frequencies drop down quickly, so the cosine is strong near the output surface of the PBG and in the x direction parallel to its face, but it drops down as well as the field propagates in the longitudinal direction: in the far field it no more survives leaving only the propagating field without evanescent components. Many 1D multilayer systems have a transmission function that may be approximated with the one described and usually with more pairs of Dirac-$\delta$'s. We will see in fact that these frequencies are strictly linked to the geometry of the system and have no dependence from the input field. These particular spatial frequencies generate at the output, cosine terms that disturb the process of super resolving. In the following we will give more details about the physical mechanism that gives rise to these peaks in the evanescent spectrum of many PBGs. A better and more realistic situation, especially if absorption is present is the one in which the structure present spectra similar to the one we have discussed but with peaks of finite high and width as shown in fig.1. The mathematical analysis is similar to the one we have done and the field at the output has two terms as in eq.(4),

$$V_o(x) = \Im^{-1}\left\{\mathrm{rect}_{2mk_0}(k_x)V_i(k_x)\right\} + \left[S(x)\cos(\alpha k_0 x)\right] * V_i(x). \qquad (5)$$

The first term is still the Fourier antitransform of the input electromagnetic field after cutting high spatial frequencies with $|k_x/k_0| > m$ (this term takes into account the loss of high spatial frequencies). The second term is a convolution between the cosines with a function $S(x)$ (that is the Fourier transform of the peaks at $\pm\alpha$ shown in the figure) which is due to the finite sizes of the peaks and gives an oscillatory symmetric function that decays with increasing $|x|$. In this case too, the peaks add a disturb to the output signal but now span only for a limited extension on the x axis.
If for example as input signal we use a rectangular function along x then the output field, due to the first term in eq.(5) will still be almost rectangular (the high spatial frequencies are cut from the transmission function) superposed with a function that oscillates from the origin reducing as $|x|$ increases. This is what happens on the output surface. If we consider the propagation of the field in the z direction, all the evanescent oscillatory components decay very soon leaving just the propagating field. It is clear from this discussion that the more are the peaks the more is the disturb at the output.
We show now an example taking into account a particular PBG. In fig.2 a metal-dielectric PBG with five layers is shown, the metal is silver while the dielectric is without absorption and with refractive index $n_d$=4. The wavelength is $\lambda = 0.600\mu m$ and for the silver we take $n_{Ag}$=0.1243+i3.7316. The polarization is TM. The thickness of all the layers is 20nm, so the whole PBG is large 100nm. The transmission and reflection spectra are shown in fig.3. The transmission spectrum has a finite band-width extending approximately from -6 to +6 $k_x/k_0$ with three broaden maxima: one in the real zone between $|k_x/k_0| < 1$ and two symmetric around $k_x/k_0 = \pm 2$. As an example of the response of the system assume to enter with a rectangular wave with a width of 100nm from the left side of the system. The spectrum of the input rectangle is plotted in fig.4a till $k_x$=40$k_0$. In fig.4b we have the anti-transform of fig.4a. Fig.4c shows a smooth approximation of the transmission spectrum of the system of fig.2 which extends up to $k_x=6k_0$ (that is the cutting level of the PBG), and in fig.4d there is the anti-transform of the truncated spectrum 4c. As we can see, for a rectangle of 100nm of width, if we truncate its spectrum at six times $k_0$ we do not introduce substantial enlarging of the spatial signal, this wave travels inside the PBG mostly without diffraction up to the output.

## 3. IS IT POSSIBLE TO ENLARGE THE $K_X$ SPECTRUM?

In this section we want to give a physical explanation of the broad peaks as the ones shown in fig.3 that are typically present in multilayer structures in the evanescent zone. We show here that these peaks may be associated to the presence of real modes propagating in the direction parallel to the multilayer planes. We start studying an ideal case, and after we will extend our results to real cases.
Without loss of generality we may analyze the PBG plotted in fig.5. The polarization is TM. The multilayer can be seen in the z-direction as an open resonator, in which some modes exist. The mathematical analysis to find the modes follows exactly the one used in optical guides in which light propagates in the x-direction; the guides indeed in the transverse direction are resonators. From the studies of the modes inside the guide we can find the dispersion curves that link the modes at the couple of values $k_x$ and $\lambda$. The dispersion curve for the system of fig.5 is shown in fig.6 (it appears discretized just due to the representation technique). In this example, below $\lambda = 0.5\mu m$ there are no more dispersion curves. In fig.6 we analyze only the window $k_x / k_0 > 1$ because the modes must be evanescent outside the PBG [11]. If we excite the PBG with a plane wave, at $\lambda = 0.5\mu m$ with a generic $k_x$ (red dashed line in fig.6), we are not able to exactly excite any mode but, because we are very close to the resonant dispersive curve at $\lambda = 0.52\mu m$ with $k_x / k_0 \approx 2$, the field inside the resonator may grow somehow and with it the evanescent tails outside the multilayer will grow. In fig.7 we show the angular transmission for $\lambda = 0.5\mu m$. For the range of values $k_x / k_0$ that are closer to the dispersion curve, say $1 < k_x / k_0 < 3$, there is a very broad transmission without peaks (there are no peaks because we are not exactly exciting modes). If we now select the wavelength at $\lambda = 0.65\mu m$ (green dashed line in fig.6) we should have a broad transmission inside the range $2 < k_x / k_0 < 5$ while a mode at $k_x / k_0 \approx 8$ could be stimulated which generates a peak in the spectrum exactly at this value (s. Fig.8). We have only to underline that the peaks are not very high just because computing discretization was low. In fig.9 we plot the dispersion curve for the system of fig.5 but for TE polarization. In this case for every wavelength there is at least one $k_x$ value; so it is impossible in this case to generate a simple broadening of the band without peaks in the spectrum. In fig.10 we show the case where $\lambda = 0.600\mu m$. We have a peak at $k_x / k_0 \approx 1.3$ that is exactly the intersection between the dashed red line and the dispersion curve of fig.9.

Let now consider the same structure of fig.5 in which now the two layers with n=i4 are substituted with two metal layers with n = 1+i4. We choose a high imaginary value just to test the goodness of the theory. After we will study a realistic metal-dielectric multilayer made with silver. The polarization is TM. If we analyze the dispersion curve we obtain the curve of fig.11. Before proceeding further, we have to explain how it was obtained and which are the differences with the dispersion curve of the previous system. From the theory of a wave guide [11] described by a 2x2 transmission matrix, we know that to find the guided modes it is necessary that the element $m_{22}(k_x,\lambda)$ of this matrix satisfies the relation

$$m_{22}(k_x, \lambda) = 0 . \tag{6}$$

From eq.(6) it is possible to find all possible modes that can travel inside the guide. For the lossless case eq.(6) has roots just because there are modes that can survive for an infinite time inside the cavity of the waveguide, but if losses are present eq.(6) cannot be satisfied (inside the cavity there are no modes that can survive for an infinite time). In this case the function $m_{22}(k_x, \lambda)$ can just be close to zero but without reaching it. The case of absence of roots for eq.(6) is exactly the situation of the PBG with losses, the real modes do not exist anymore but there are points of resonance (the minimums of the function $|m_{22}(k_x, \lambda)|$) where we can increase the field inside the cavity at values much higher than the input. In fig.11 we plotted all the minima of the function $|m_{22}(k_x, \lambda)|$. It is clear that the larger are the losses of the system the more distant are the minima from zero and the weaker are the corresponding resonances. In fig.12 we plot the angular transmission at $\lambda = 0.6\mu m$. In this case the dashed line at $\lambda = 0.6\mu m$ in fig.11 intersect the dispersion curve but as we can see from fig.12 the transmission grows keeping it limited and generating a smoothed peak exactly in the intersection point $(\lambda = 0.6\mu m,\ k_x / k_0 = 1.28)$. This limited and smoothed peak at intersection is due to the fact we are no more in presence of a real mode but just of a resonance, where the losses limit the field amplitude inside and outside the cavity. In fig.13 the dispersion curve for the same system is plotted for TE polarization. In this case too we always intersect the dispersion curves, but here too due to losses we have a broadening enlarging the transmission spectrum without peaks. This is seen in fig.14 in which the transmission function for $\lambda = 0.6\mu m$ is plotted. A band with the maximum at $k_x / k_0 \approx 1.21$ that corresponds to the intersection between the dashed line with the dispersion curve, fig.13 is seen.

The conclusion is that the broad peaks that appear in the spatial spectrum of the multilayer structure for $|k_x / k_0| > 1$ correspond to the evanescent tail of transverse wave propagating parallel to the multilayer planes which can be excited from outside.

## 4. DESIGN RULES FOR LARGE BAND PBG

In this section we want to discuss some ideas for designing metal-dielectric 1D-PBG with large band and good transmission over the entire spectral range. We analyze for the moment only symmetric systems as the one plotted in fig.15. All the dielectric layers have the same thickness $d_d$ and refractive index $n_d$, all the metal layers have the same thickness $d_m$ and refractive index $n_m$. In the analysis we have to search a geometry that is able to enlarge the band without generating peaks in the spectrum. First we fix the wavelength at the value $\lambda_0$ where the dielectric refractive index and the imaginary part of the metal refractive index are equal (or in any case very close to each other), so we need that

$$n_d(\lambda_0) = \mathrm{Im}\{n_m(\lambda_0)\}, \tag{7}$$

and we choose a metal where the real part of the refractive index is very low with respect to the imaginary one,

$$\mathrm{Re}\{n_m(\lambda_0)\} << \mathrm{Im}\{n_m(\lambda_0)\}. \tag{8}$$

An ideal metal for which $\mathrm{Re}\{n_m(\lambda_0)\} = 0$ would be the best choice, but if the real part of the metal refractive index is very low compared with the imaginary one, all the discussion in the following is still valid. The reasons for these choices will be explained later.

Now we apply rules (7) and (8) to a real case. We choose $\lambda = 388\mathrm{nm}$, in this region the silver has a refractive index $n_m = 0.1824 + i \cdot 1.8164$. Therefore we choose a dielectric with refractive index around 1.8 at $\lambda = 388\mathrm{nm}$. We choose six silver and seven dielectric layers to build our PBG. To start with we fix all the thicknesses of the layers equal to each other and rather large, for example we can start with $d_d$=$d_m$=60nm and then we decrease the thickness of the layers to see which is their influence on the transmission spectrum. Several cases are shown in fig.16, which shows the angular spectrum for different values of the thickness. We see that, with the large thickness we have chosen to start with, the angular spectrum of the system is regular (without resonances), narrow and with a very low transmission value (fig.16a). Reducing the thickness of the layers, the spectrum starts to open. For $d_d$=$d_m$=30nm (fig.16b) and $d_d$=$d_m$=20nm (fig.16c), the spectrum is rather broadened. For $d_d$=$d_m$=15nm (fig.16d) we see that the modulus of the propagating waves reaches a maximum value of about 15% in transmission while the evanescent components still grow. For smaller thicknesses we increase the resonances of the evanescent field but the transmission spectrum is no more uniform. So we may decide to stop the thickness of the layer at about 20nm just diversifying a little bit the thickness of the metal from the dielectric to find a more uniform transmission. We choose 18nm for the dielectrics and 20nm for the metal and finally we have the spectrum of fig.17a which has a low modulus and a spectral width larger than four. In fig.17b the dispersion curve is plotted. The dashed line ($\lambda = 0.388\mu m$) intersect the curve at about $k_x / k_0 = 2$ where there is the maximum of the transmission, the peaks are not infinite just because there is the absorption of the silver.

Now we show how important is to have the condition (8) well verified. We modify the refractive index of the silver to $n_m = 0 + i \cdot 1.8164$. The spectrum of fig.17a modifies in the one of fig.18a showing a very large width which demonstrates that this system with ideal metal and symmetric geometry is able to avoid the resonance peaks and to super resolve a signal large at minimum 1/10 of the wavelength. From this example we see that it is not important to have a very large refractive index for the dielectric to have a large band in the spectrum but it is important that eqs.(7) and (8) are well verified. In fig.18b the dispersion curves are plotted and one of them is tangent to the line $\lambda = 0.388\mu m$ (dashed line); just because the discretization is low the dispersion curve up to the line $\lambda = 0.388\mu m$ is plotted with a few of points and the tangent curve appears broken, but it is continue. In fig.18c the dispersion curves are plotted always for the same system but changing the real part of the silver refractive index. We can see that just for completely absence of absorption (fig.18b) the dispersion curve is tangent to the line $\lambda = 0.388\mu m$, for all other cases we cannot avoid the intersection with almost one dispersion curve. This characteristic is completely general, when absorption is present, it is never possible to avoid the intersection between the line $\lambda = \text{cost}$ and the dispersion curves.

We analyzed many others structures using different types of systems, both symmetric or non-symmetric, and also studying all-dielectric PBGs. We observed that in the case of all-dielectric systems the polarization that better works to broaden the band is TE while usually for TM there are strong peaks close the points $k_x / k_0 = \pm 1$. In the case of all-dielectric PBG the maximum width we may have in the band is given from the maximum refractive index difference between the layers. Moreover we may have a nearly uniform spectrum just if we have some losses, the intersection between $\lambda = \text{cost}$ and the dispersive curves are indeed impossible to avoid in any case.

As an other example we consider a periodic PBG using real dielectric materials AlAs and GaAs. The entrance layer is AlAs, the periods are five and the thickness of the layers are all the same, 30nm. The dispersion curve is plotted in fig.19b and it is for the TE case. We plotted in figs.19a and 19c the transmission spectrum for two different wavelengths corresponding to the dashed lines in fig.19b. The average amplitudes of the spectra are not so high but we must remember that the contrast between the two refractive indexes of AlAs and GaAs is not so high too.

The last example we propose is a completely dielectric structure with a TM input plane wave. The structure is a periodic gap with periodicity $2\mu\text{m}$ in a germanium layer (thickness 200nm so to be completely opaque for the input wave); the distance between near gaps is 200nm (s. fig.20). The angular spectrum of the PBG at $\lambda = 600\text{nm}$ and without the Ge layer is shown in fig.21a. It is large enough to super resolve an input signal large 200nm at $\lambda = 600\text{nm}$, the peaks are smoothed by the absorption of the layers. In fig.21b real part of the Poynting vector is shown for the whole system described in fig.20, while in fig21c there is the real part of the Poynting vector with vacuum instead of the PBG. We can therefore compare the propagation with and without the PBG after the opaque layer of Ge. Simple inspection shows that in presence of the multilayer super-resolution is achieved, because the width of the propagating rays is almost constant when the PBG is present.

## 5. CONCLUSIONS

From the presented discussion one may infer that the super resolution is a phenomenon that can be completely explained as a resonant effect. It is due to the resonances inside the PBG which in the propagation direction may be seen as an open cavity. Considering the resonator representation with the resonant points, we may explain why there are often strong peaks in the angular spectrum and only in the evanescent zone, we may also easily explain the phenomenon of the uniform band broadening, linking it to the dispersion curves and the quasi resonant regimes inside the PBG cavity.

We also showed that is possible to have nearly uniform large bands for every kind of polarization and every kind of 1D multilayer structure with metal-dielectric or all-dielectric layers. We demonstrated that to have very large band it is necessary to respect as better as it is possible the conditions (7) and (8), without having very large refractive index in the layers.

Aknowledgements

The authors wish to aknowledge discussion with M.Scalora and C.Sibilia.

## CAPTION FOR FIGURES

**Fig.1:** Mathematical shape for a transmission that has to super resolve a generic input field with the introduction of two finite size peaks to simulate the real shape of the spectrum of many PBGs

**Fig.2:** Special case of a metal-dielectric PBG with five layers, the metal is silver while the dielectric is without absorption and with refractive index $n_d$=4.

**Fig.3:** Plot of the transmission and reflection spectra for the system of fig.2

**Fig.4:** a) spectrum of an input rectangle at the input front of the PBG of fig.2 until $k_x$=40$k_0$. b) anti-transform of a. c) spectrum of an output rectangle at the cutting level of the PBG of fig.2 ($-6k_0 < k_x < +6k_0$). d) anti-transform of the truncated spectrum c.

**Fig.5:** A lossless and no dispersive metal-dielectric PBG with five layers.

**Fig.6:** Dispersion curve for the system of fig.5 for TM polarization.

**Fig.7:** Angular transmission for the system of fig.5 at $\lambda = 0.5\mu m$, TM polarization.

**Fig.8:** Angular transmission for the system of fig.5 at $\lambda = 0.65\mu m$, TM polarization.

**Fig.9:** Dispersion curve for the system of fig.5 for TE polarization.

**Fig.10:** Angular transmission for the system of fig.5 at $\lambda = 0.6\mu m$, TE polarization.

**Fig.11:** Dispersion curve for the system of fig.5 substituting the materials n=i4 with others having n=1+i4, TM polarization.

**Fig.12:** Angular transmission for the system of fig.5 at $\lambda = 0.6\mu m$ we substituted the materials n=i4 with others having n=1+i4, TM polarization.

**Fig.13:** Dispersion curve for the system of fig.5 substituting the materials n=i4 with others having n=1+i4, TE polarization.

**Fig.14:** Angular transmission for the system of fig.5 at $\lambda = 0.6\mu m$, we substituted the refractive index of the metal n=i4 with n=1+i4, TE polarization.

**Fig.15:** Metal-dielectric symmetric system, the metal is silver. The number of the metal layers is six while for dielectric there are seven layers. All the dielectric layers are equal, all the silver layers have the same thickness $d_m$.

**Fig.16:** Angular transmission spectrum for the system of fig.15 at $\lambda = 388nm$, $n_d$=1.8 and a) $d_d$=$d_m$=60nm, b) $d_d$=$d_m$=30nm, c) $d_d$=$d_m$=20nm, d) $d_d$=$d_m$=15nm.

**Fig.17:** a) angular transmission spectrum for the system of fig.15 where $d_d$=18nm, $d_m$=20nm, $n_d$=1.8, $\lambda = 388nm$. b) dispersion curve.

**Fig.18:** a) angular transmission spectrum for the system of fig.15 where $d_d$=18nm, $d_m$=20nm, $n_d$=1.8, $n_m = 0 + i \cdot 1.8164$, $\lambda = 388\text{nm}$. b) dispersion curve. c) dispersion curves calculated for different values of the real part of the silver refractive index.

**Fig.19:** Analysis of periodic PBG using AlAs and GaAs, the entrance layer is AlAs, the periods are five and the thickness of the layers are all of 30nm. b) dispersion curve for TE polarization. a) transmission spectrum at $\lambda = 0.5\mu\text{m}$. c) transmission spectrum at $\lambda = 0.47\mu\text{m}$.

**Fig.20:** Multilayer system with artificial and absorbing materials, at the input there is a periodic germanium gap layer that is completely opaque for the input plane wave that has $\lambda = 600\text{nm}$.

**Fig.21:** a) angular spectrum of the multilayer of fig.20 without the germanium gaps layer at $\lambda = 600\text{nm}$. b) real part of the Poynting vector for the system described in fig.20. In input there is a TM plane wave with $\lambda = 600\text{nm}$. c) real part of the Poynting vector for the system described in fig.20 but with the vacuum instead of the PBG, in input there is a TM plane wave with $\lambda = 600\text{nm}$. All figure are in arbitrary unit.

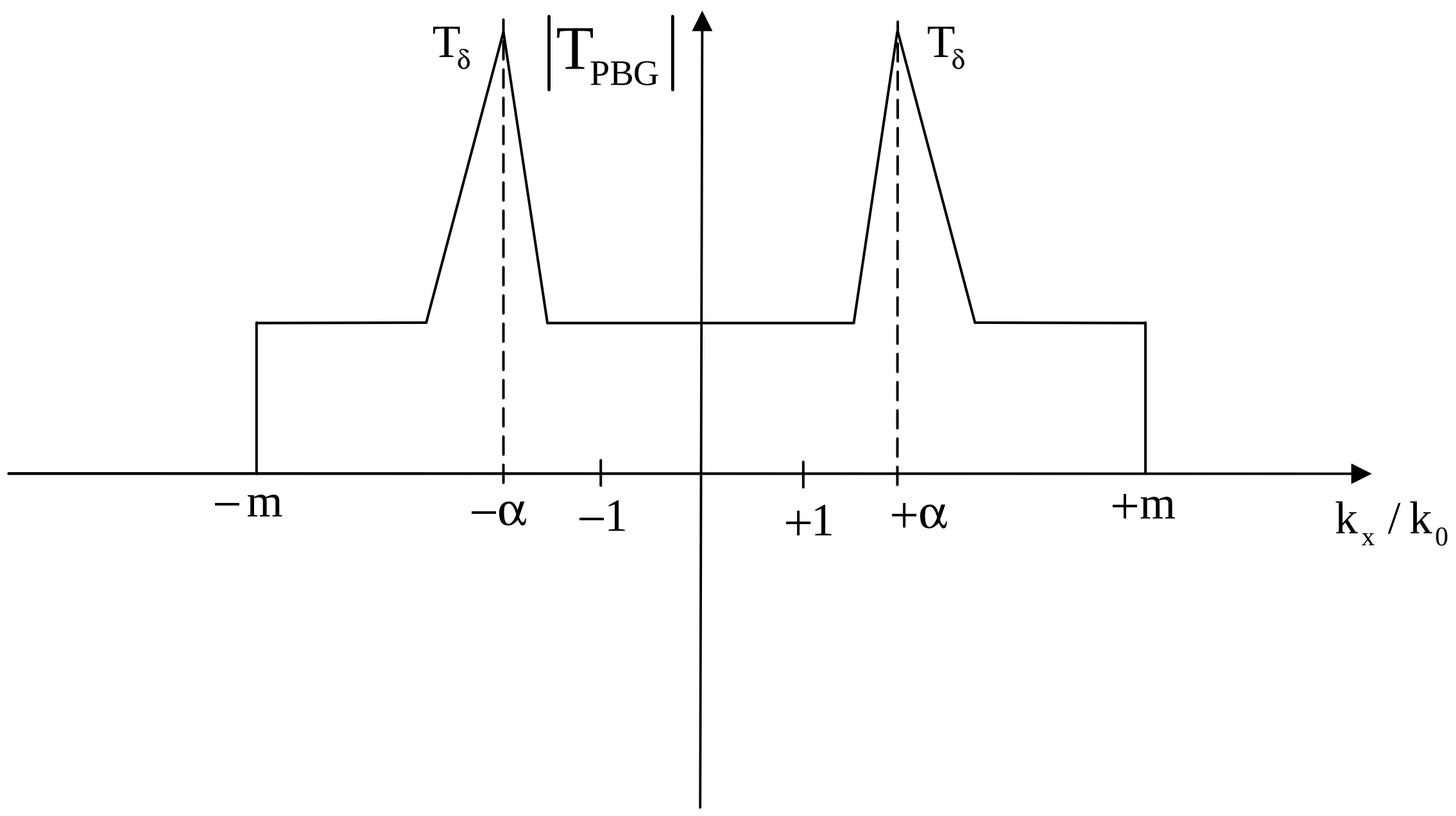

Fig.1

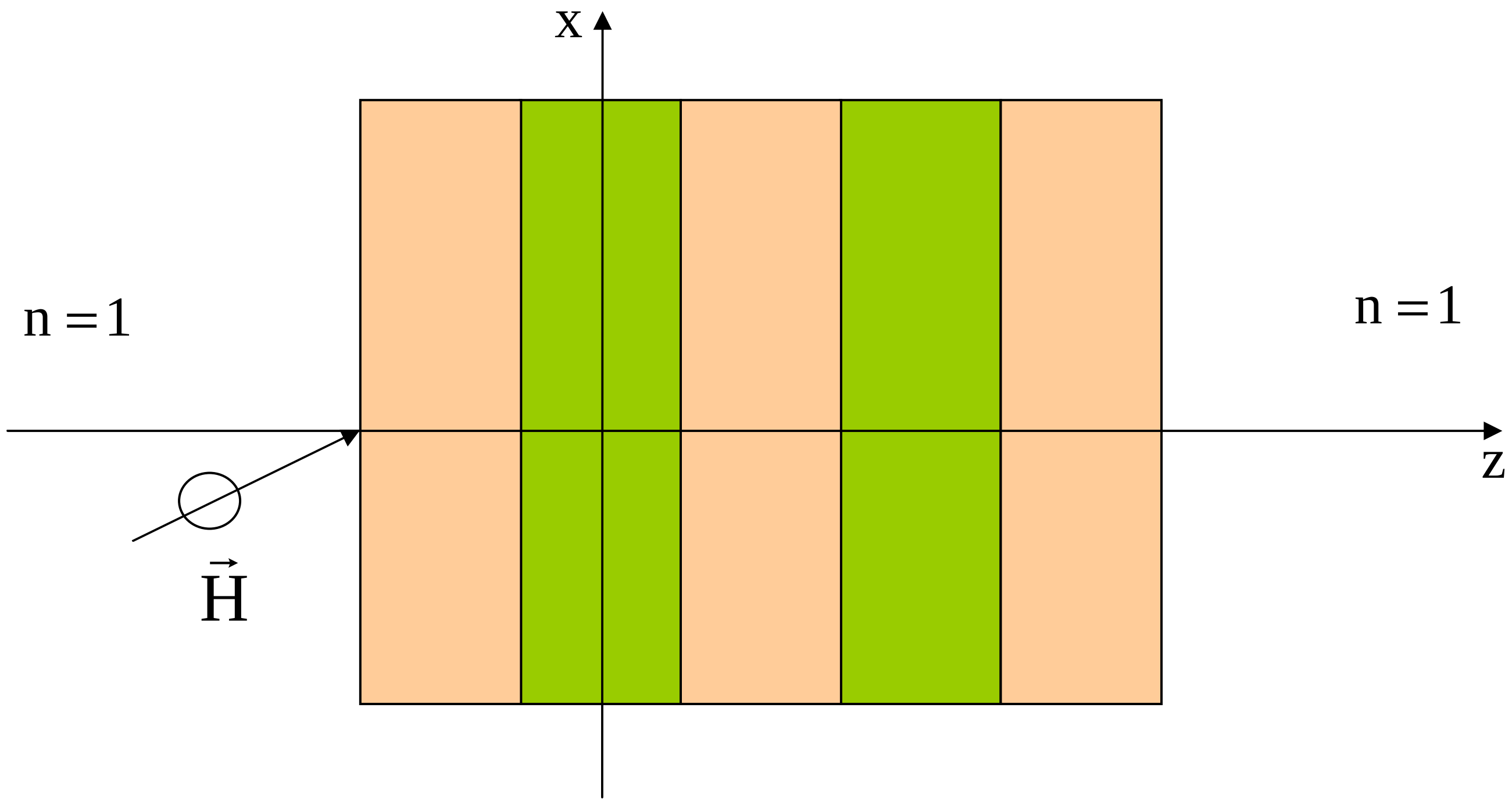

Fig.2

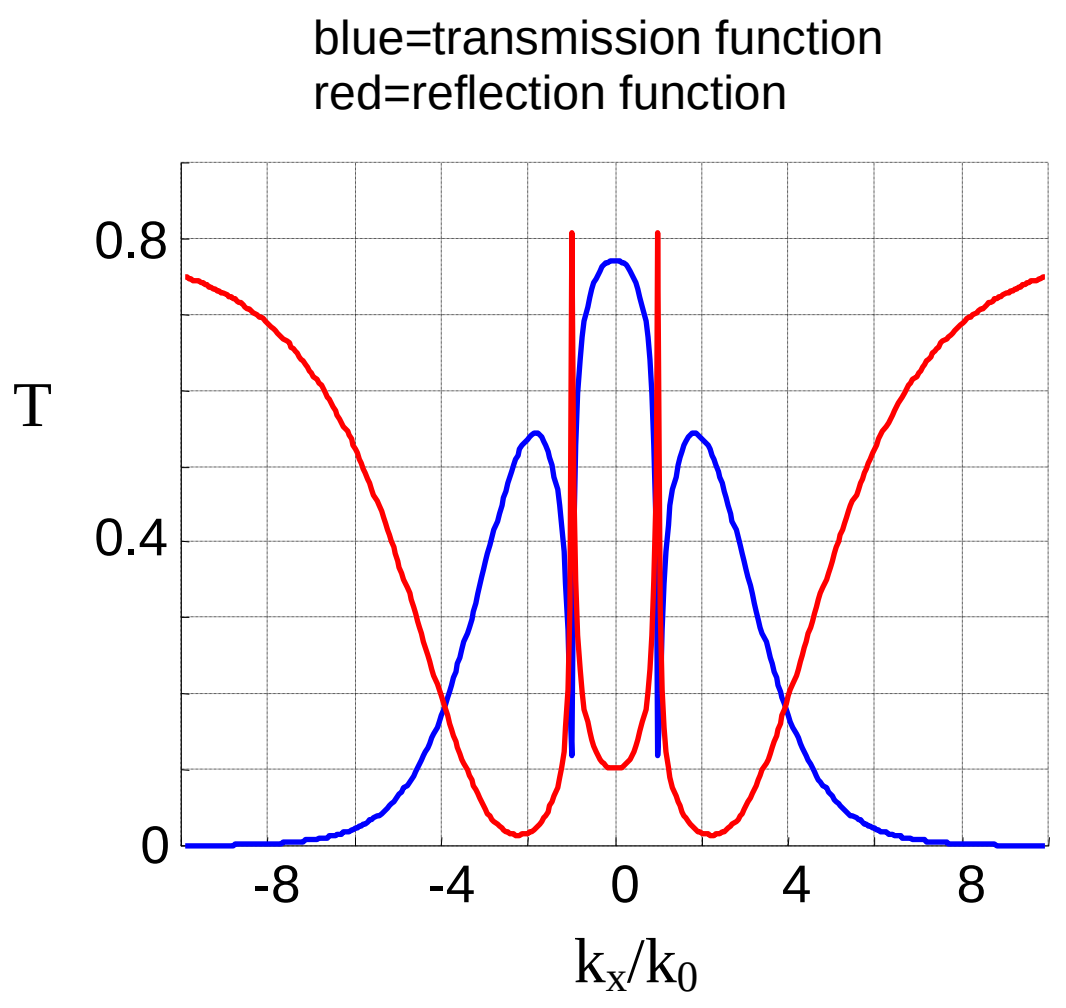

Fig.3

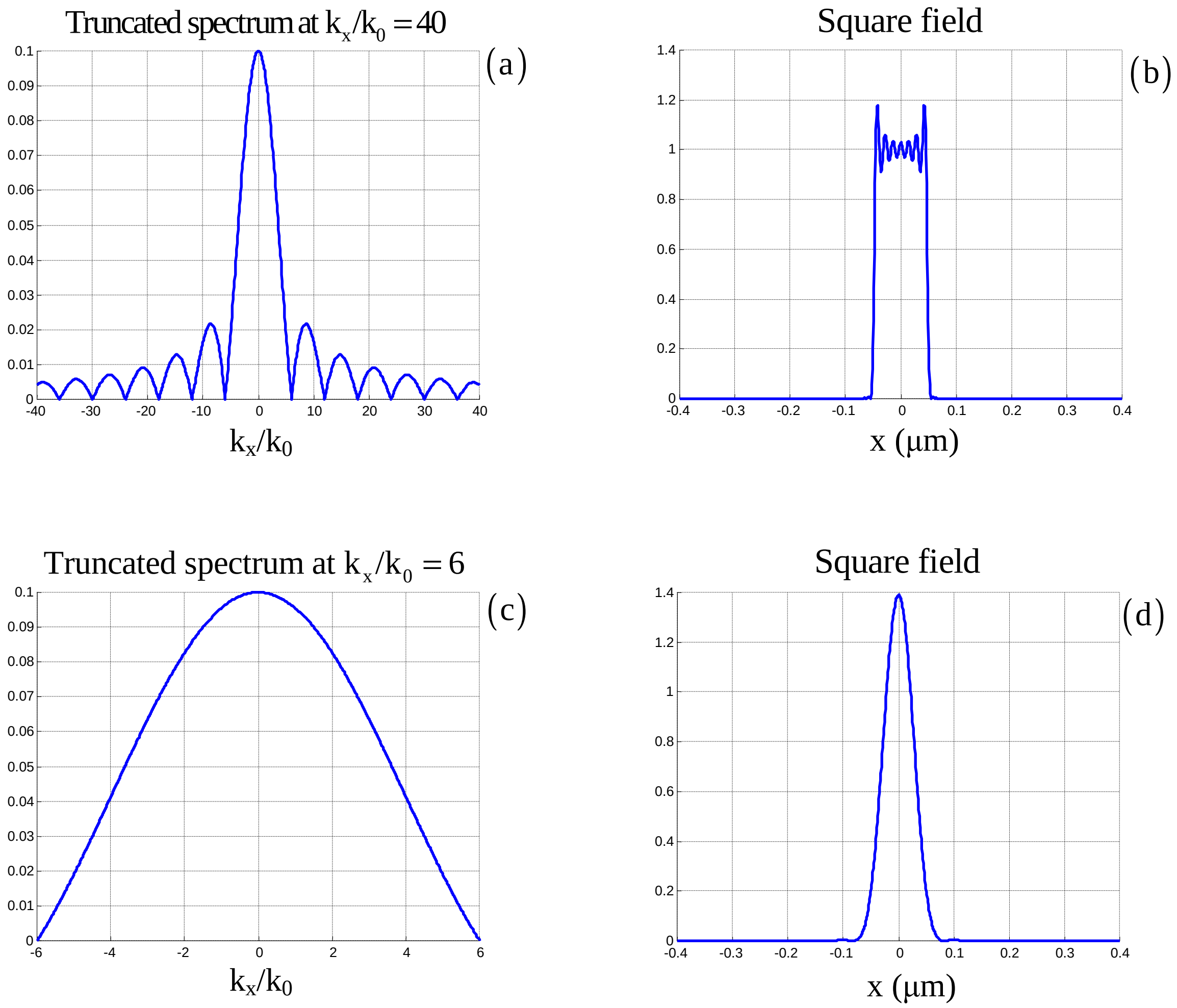

Fig.4

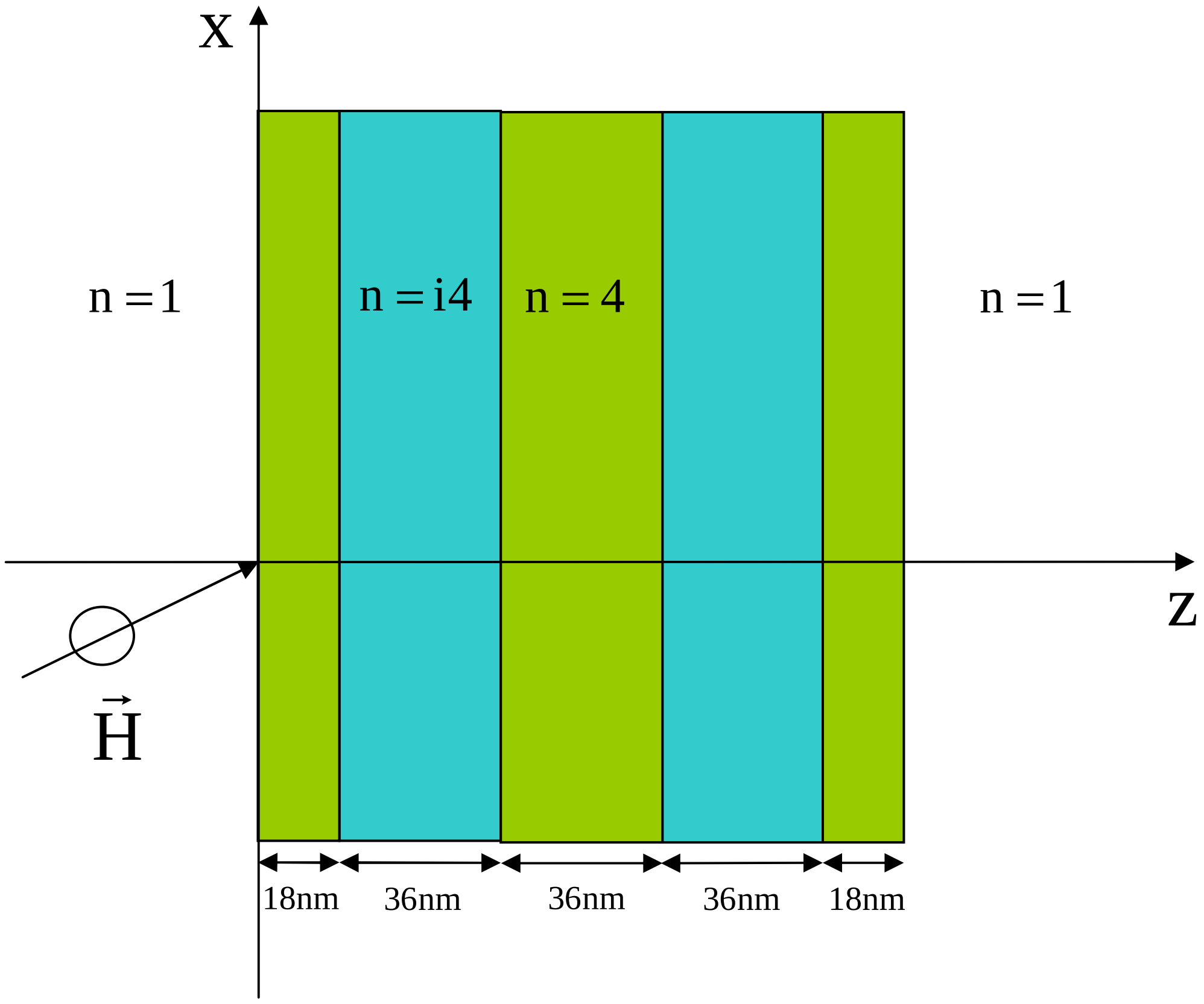

Fig.5

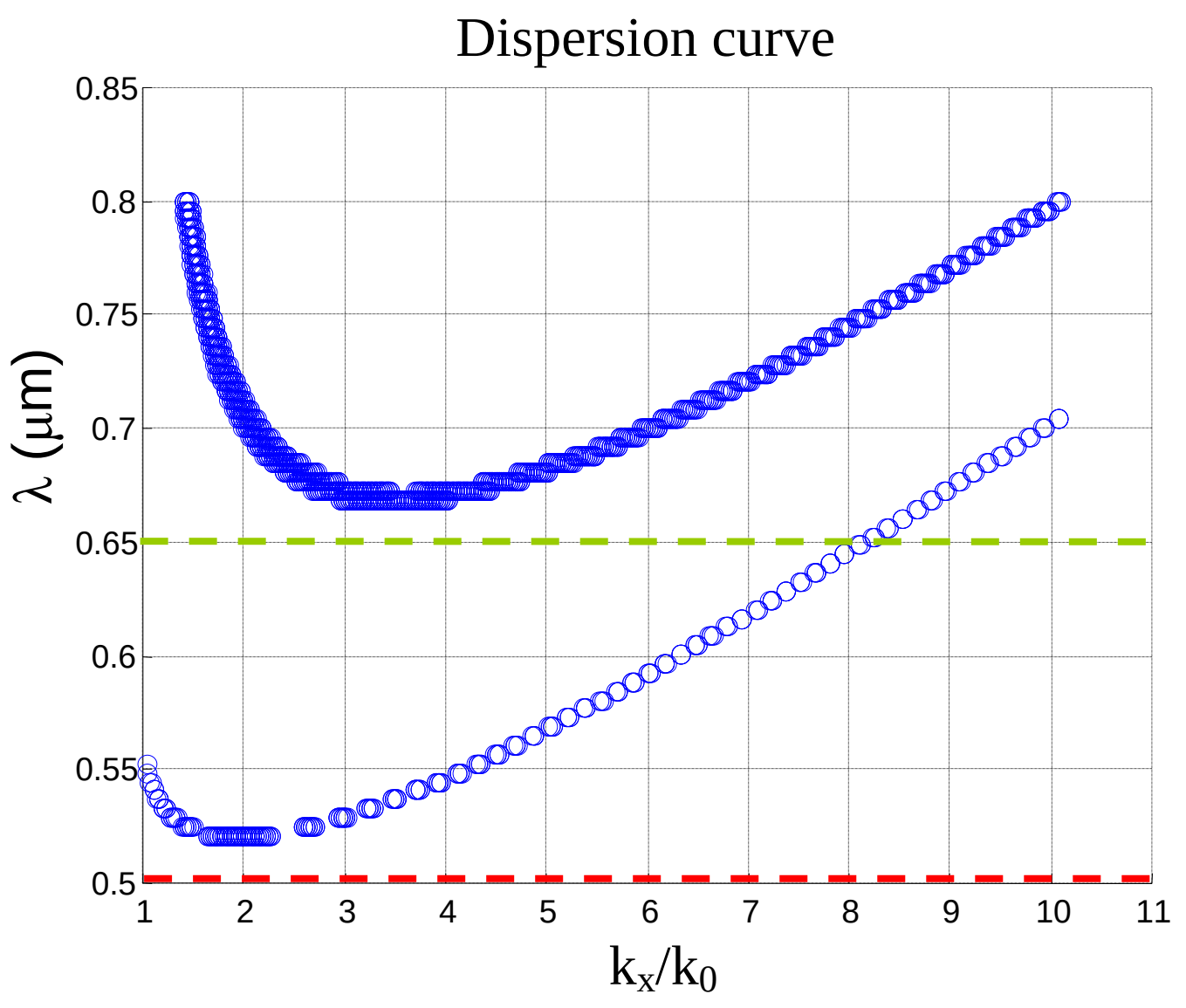

Fig.6

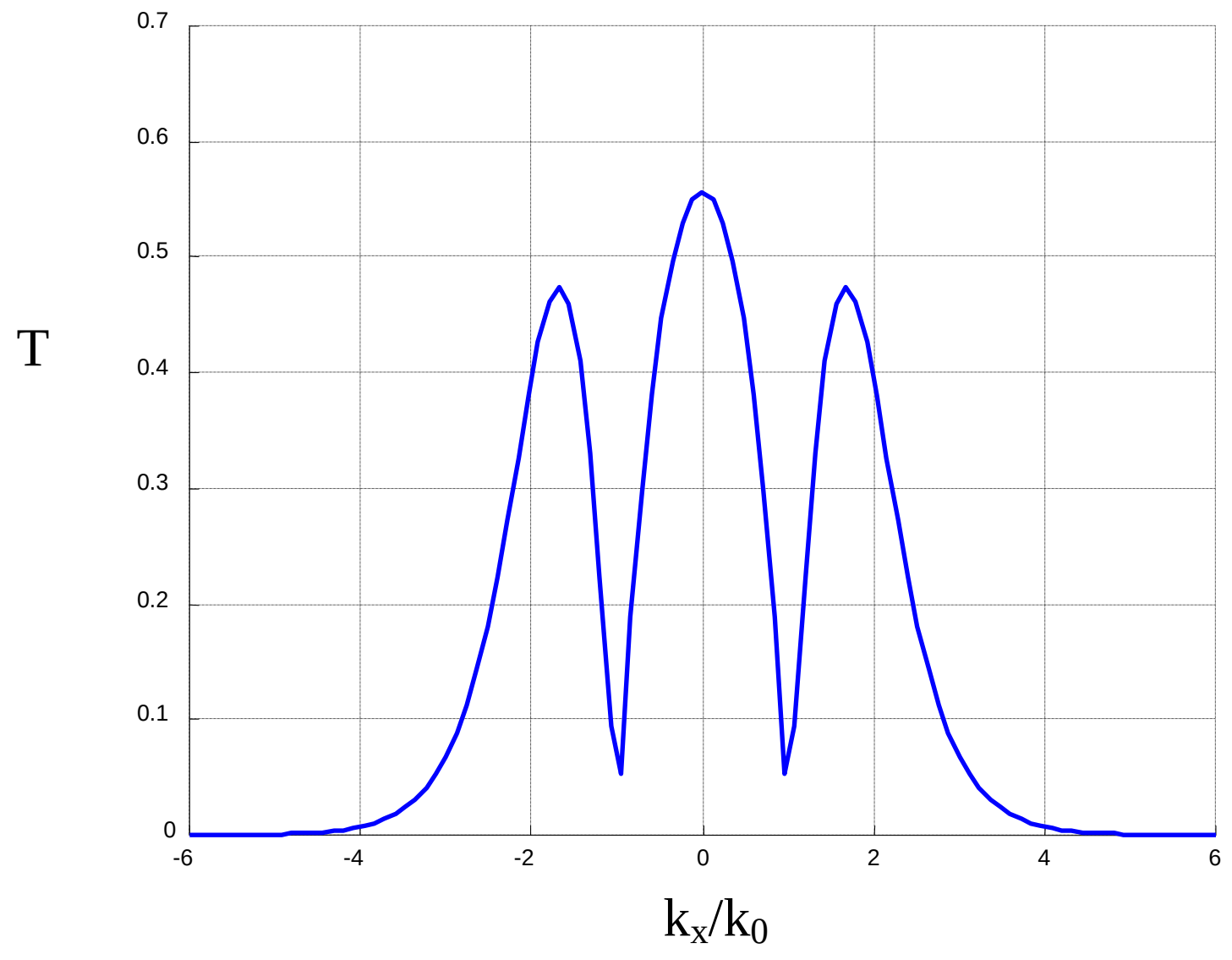

Fig.7

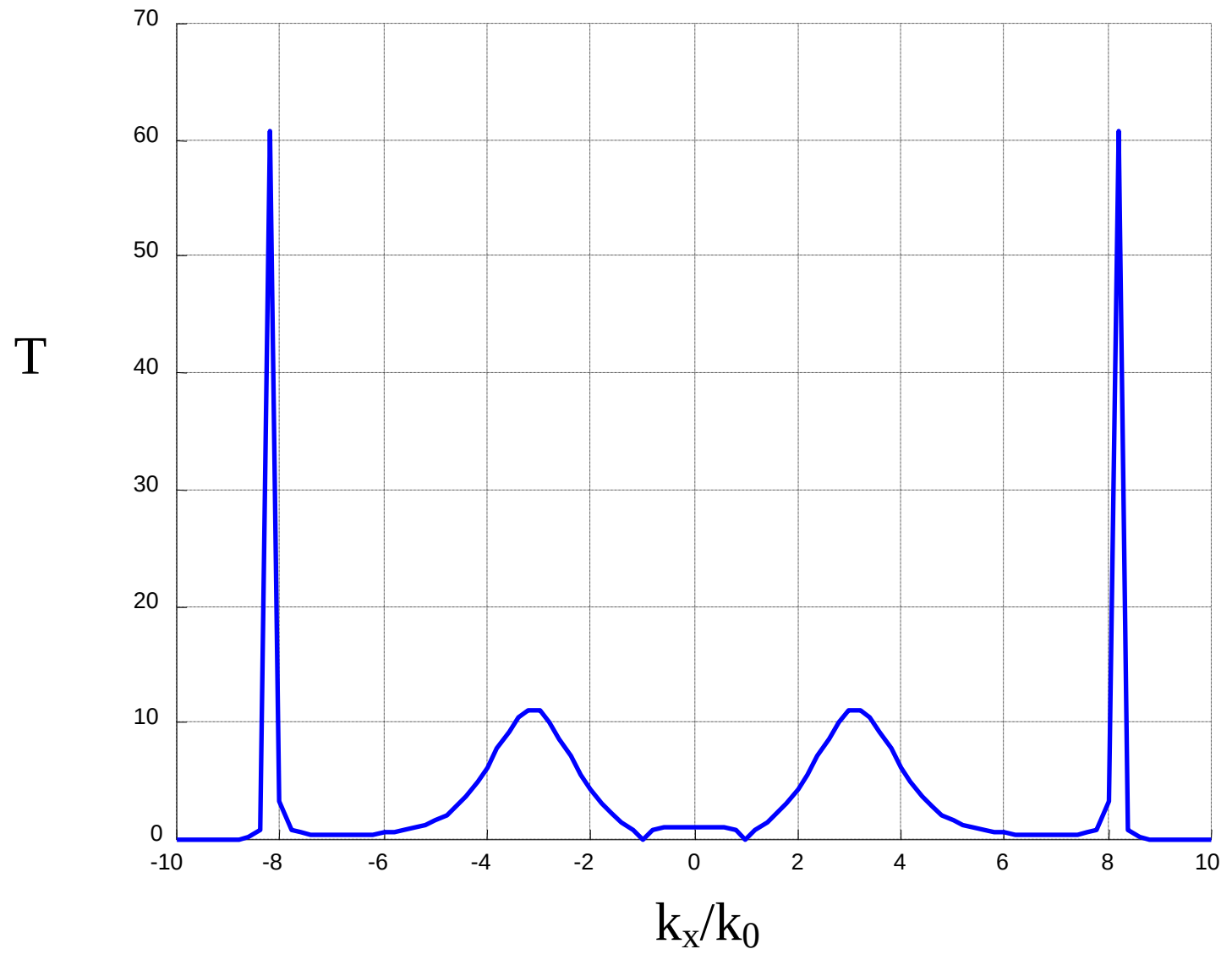

Fig.8

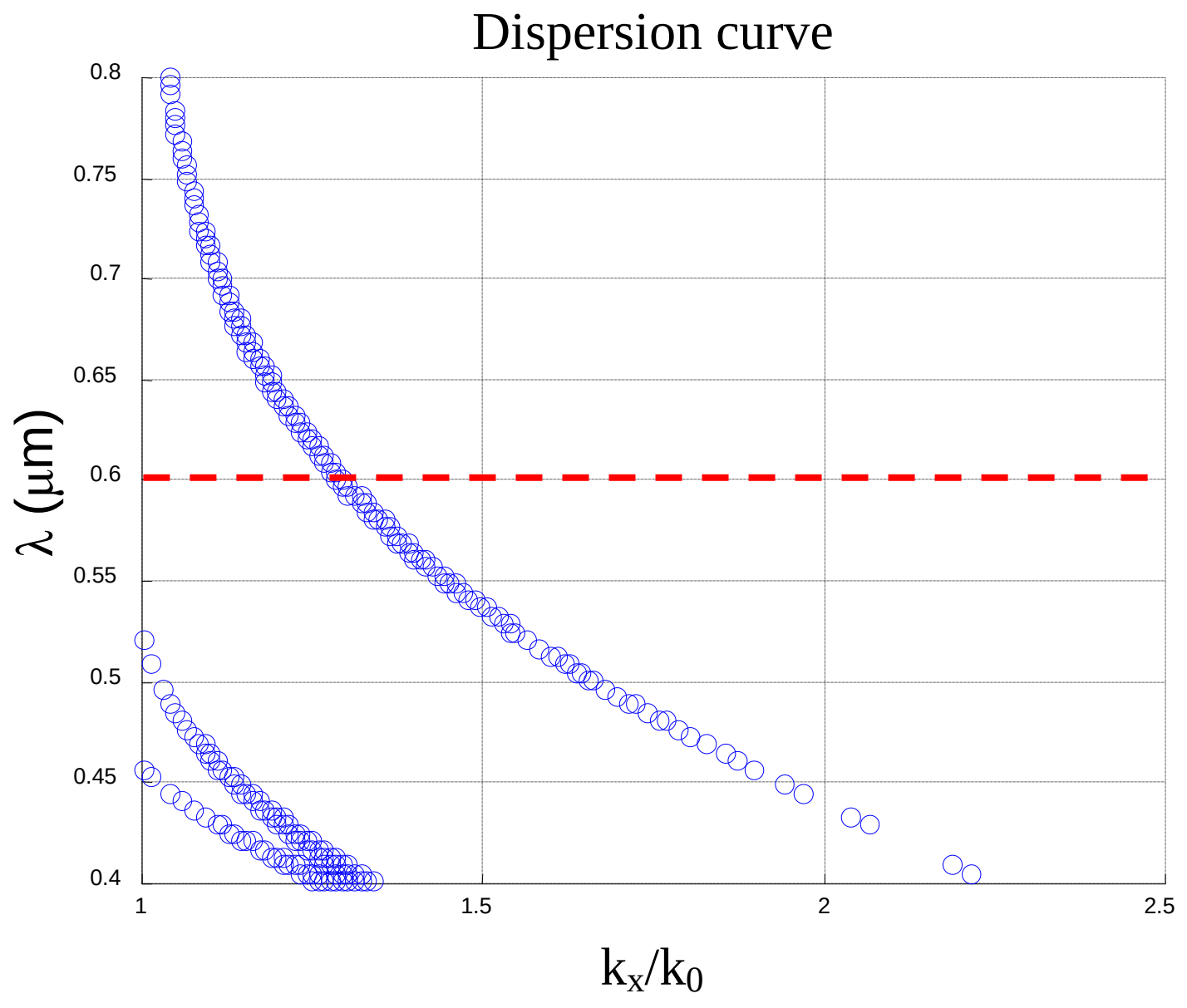

Fig.9

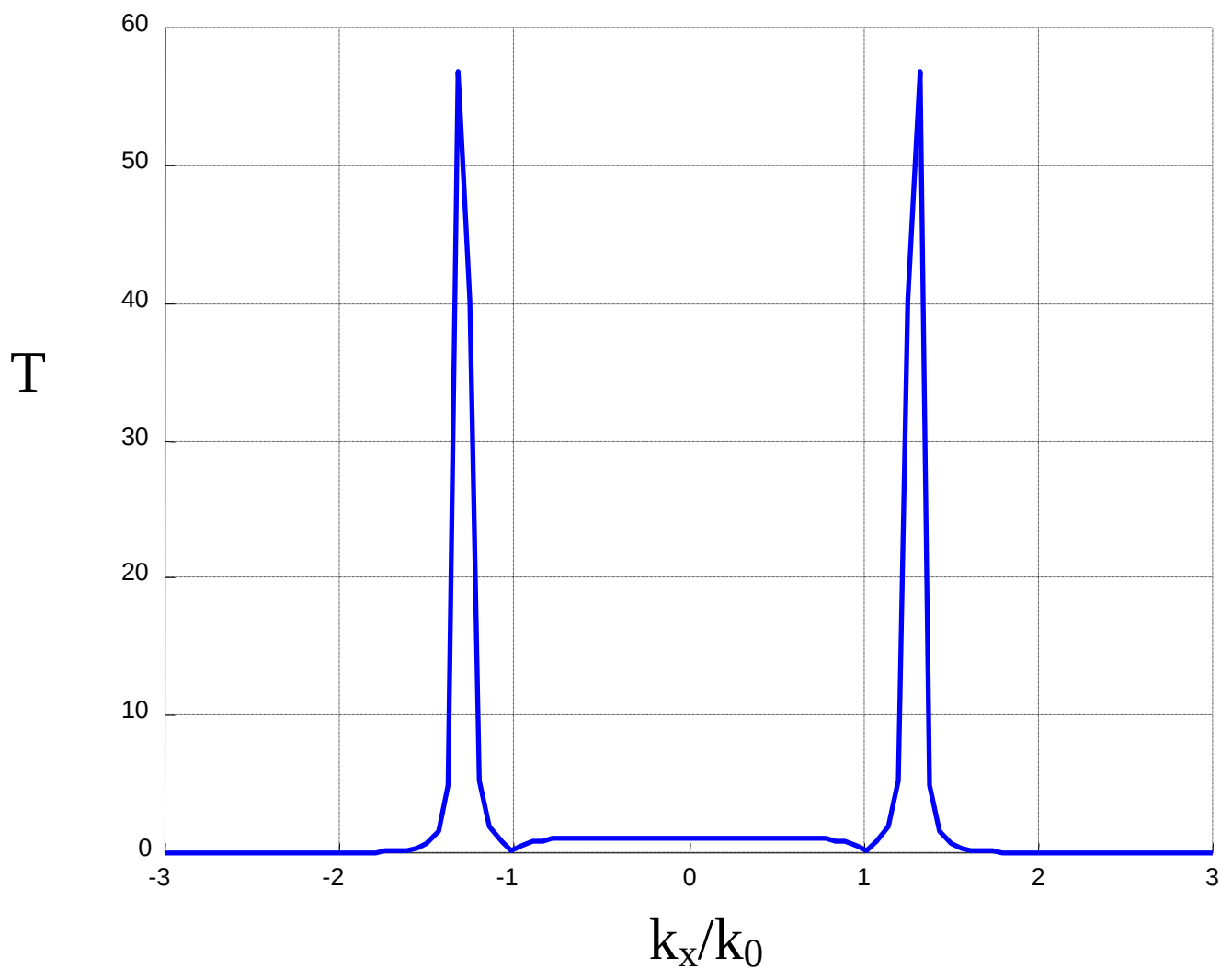

Fig.10

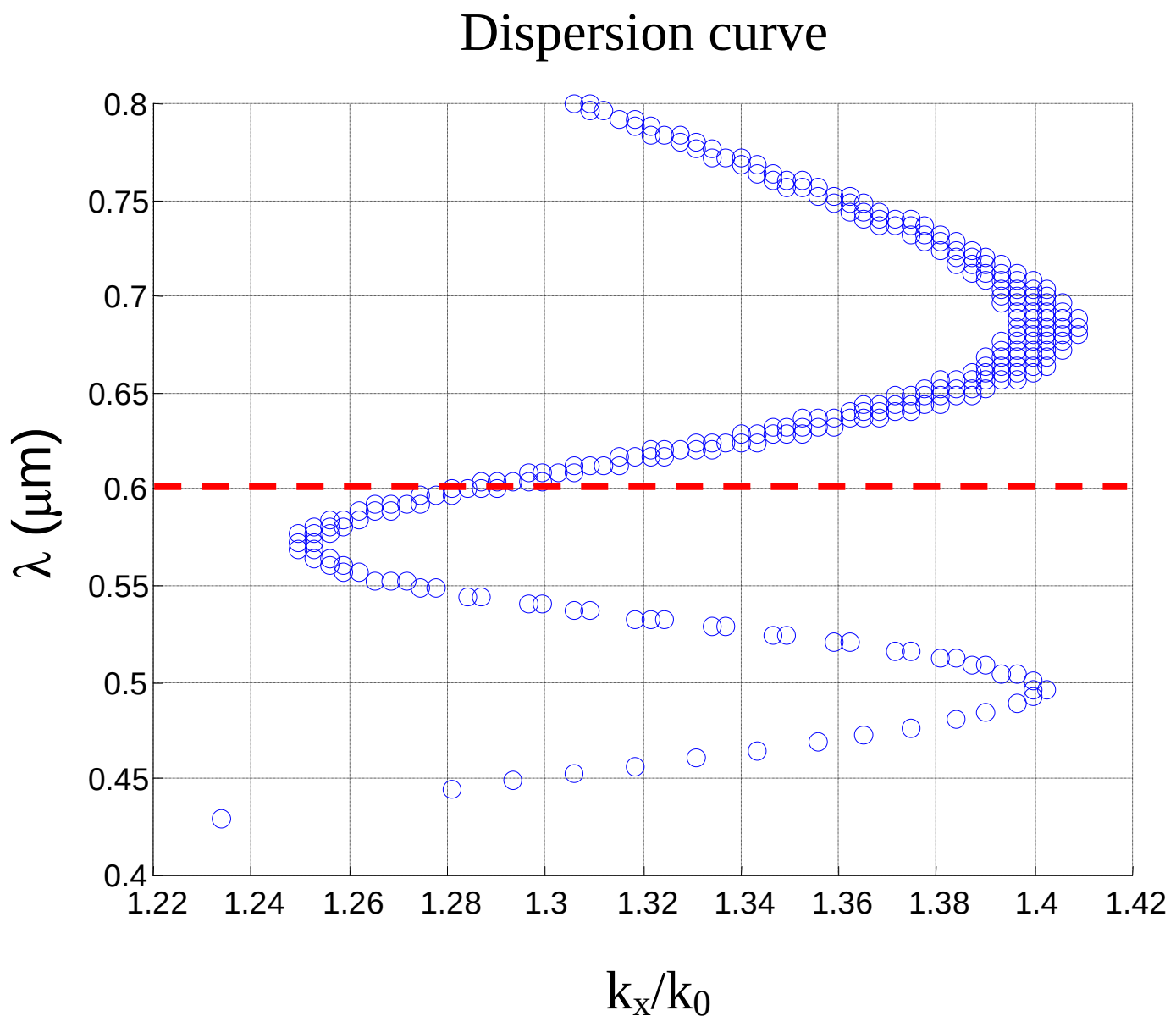

Fig.11

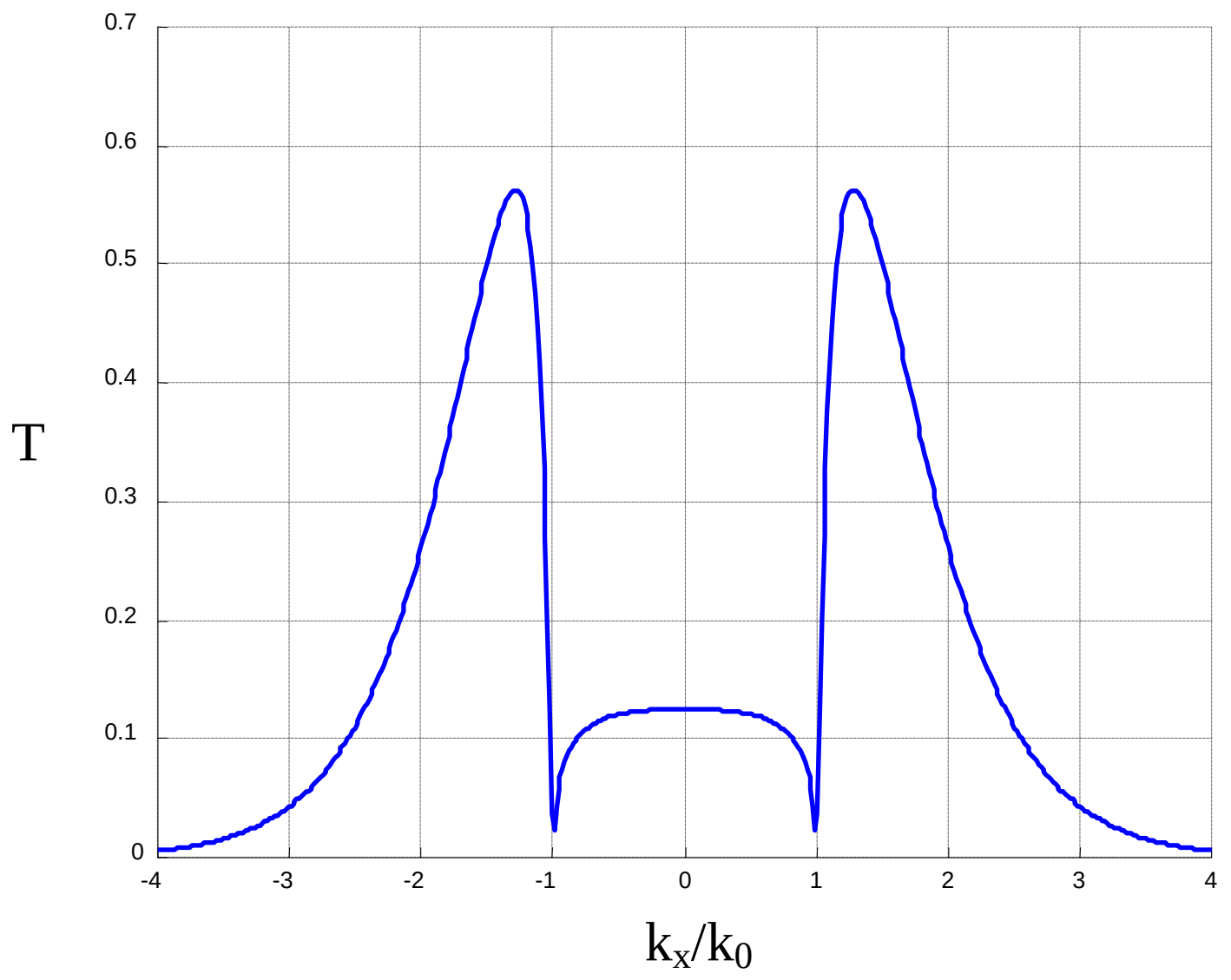

Fig.12

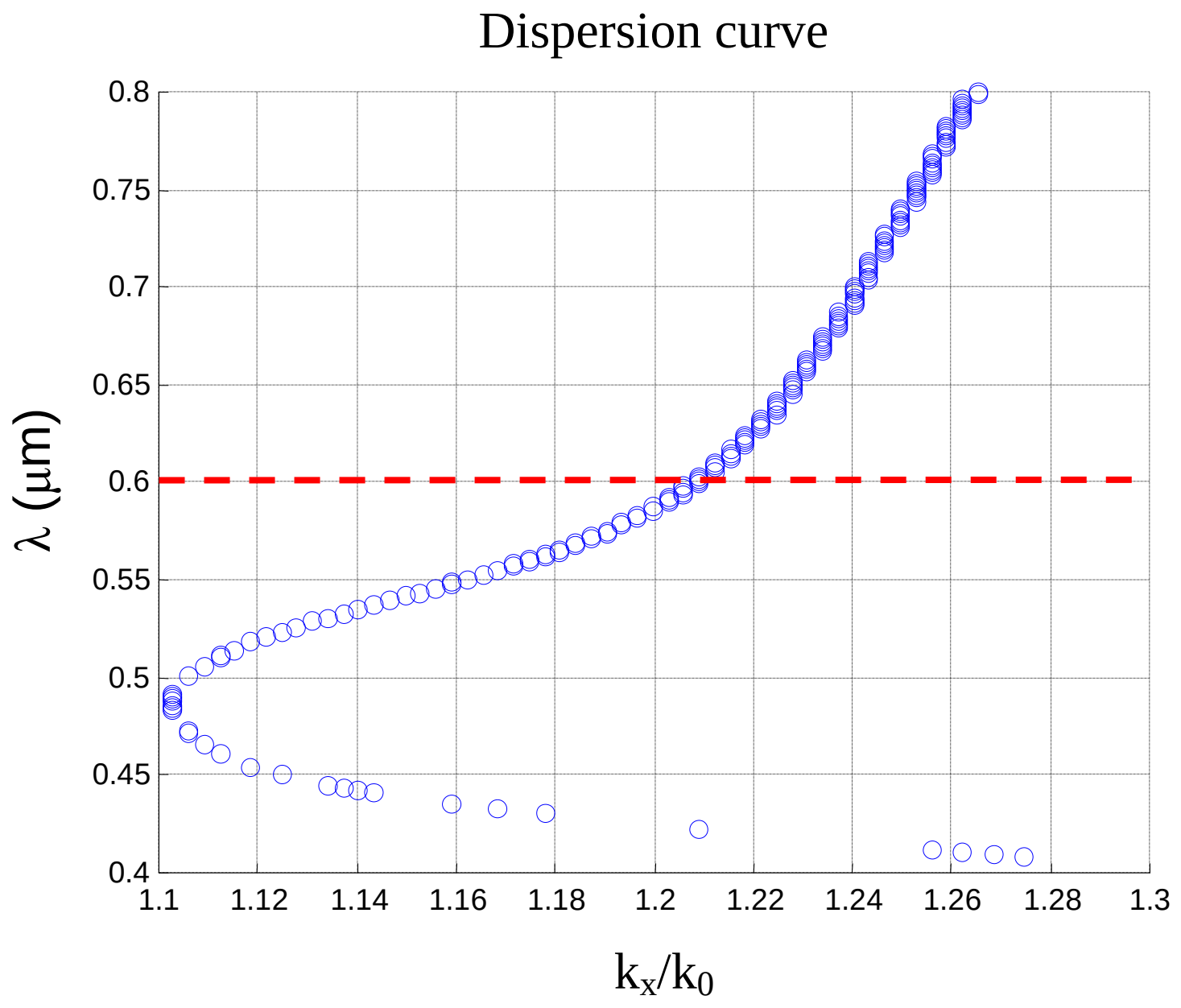

Fig.13

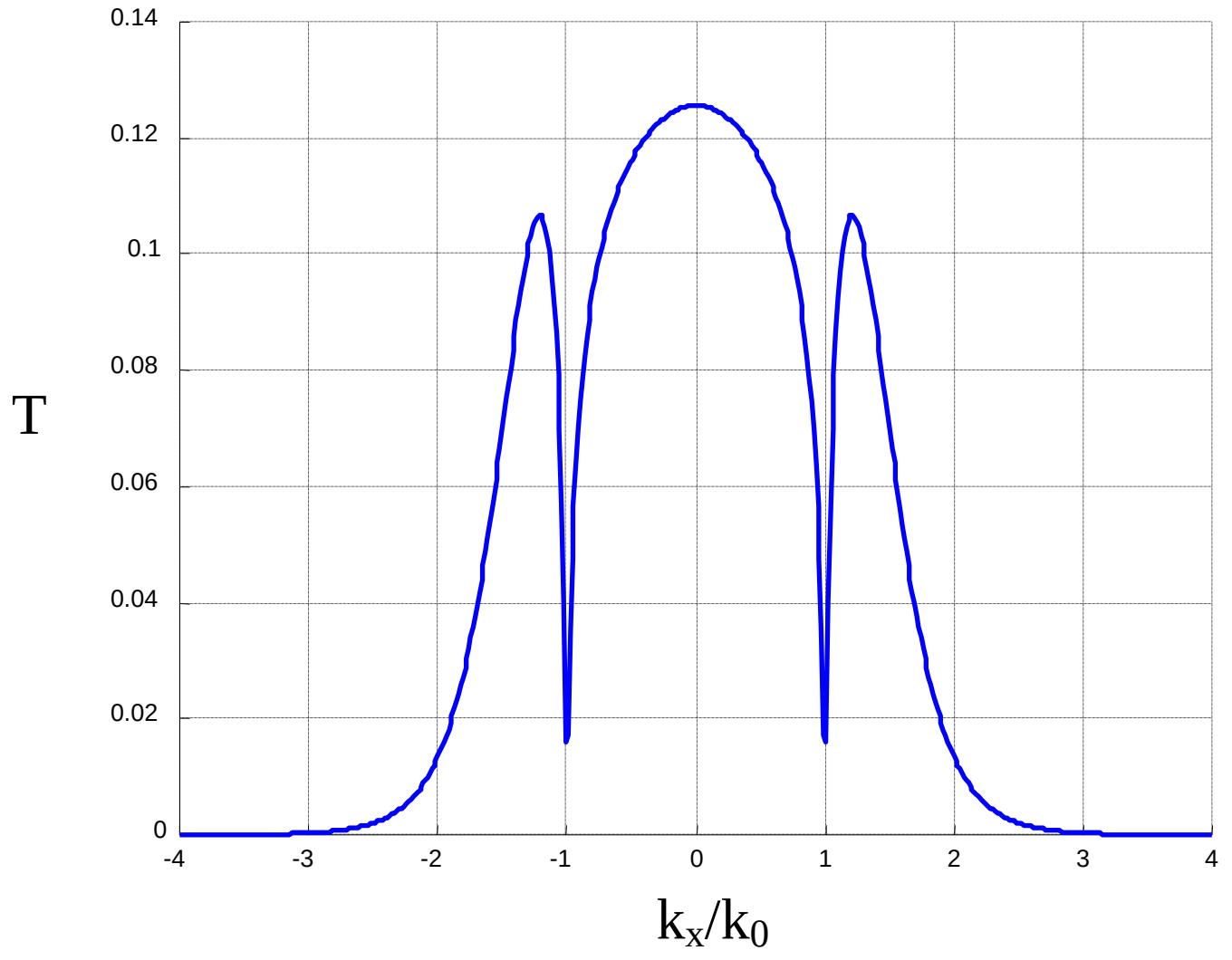

Fig.14

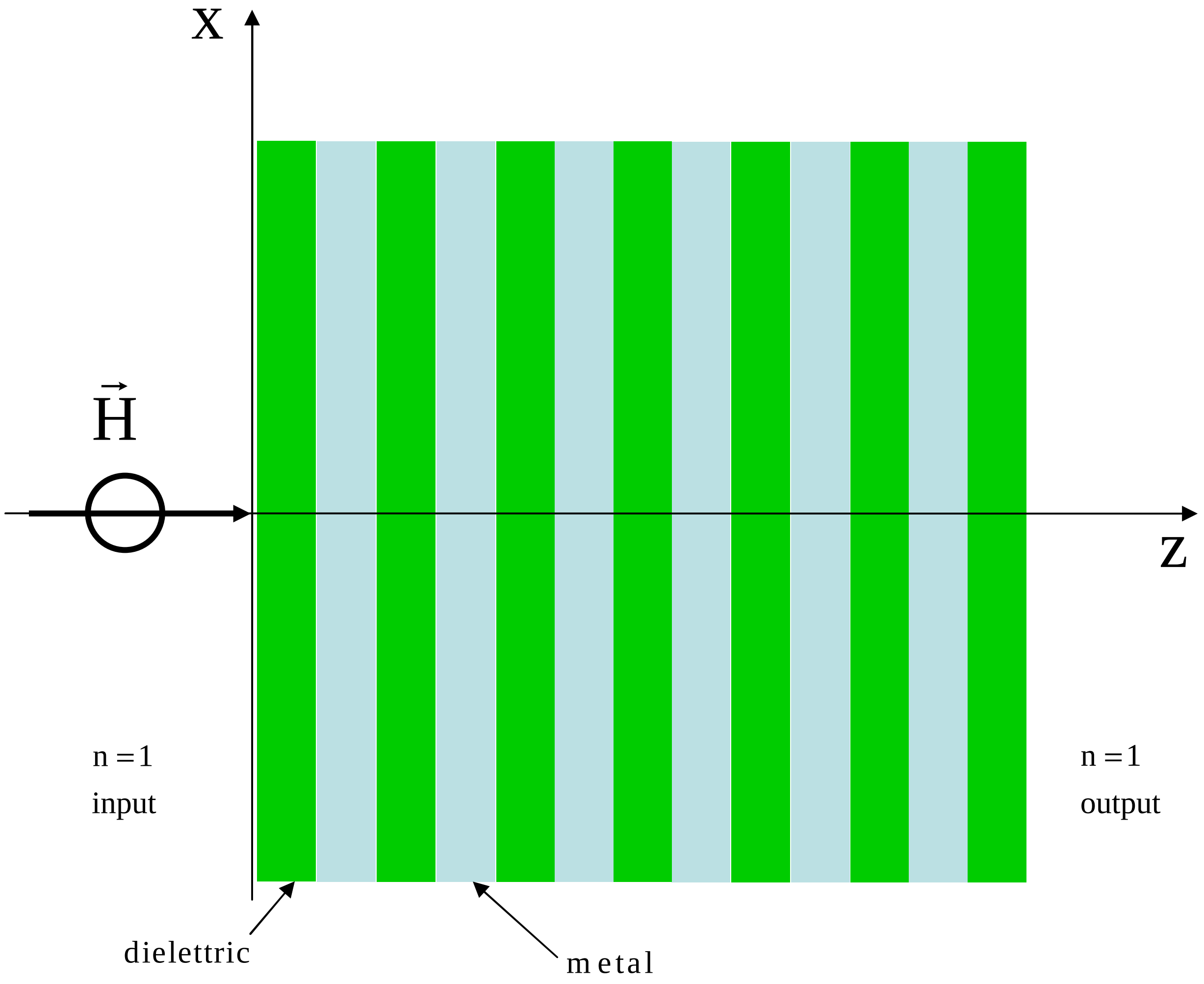

Fig.15

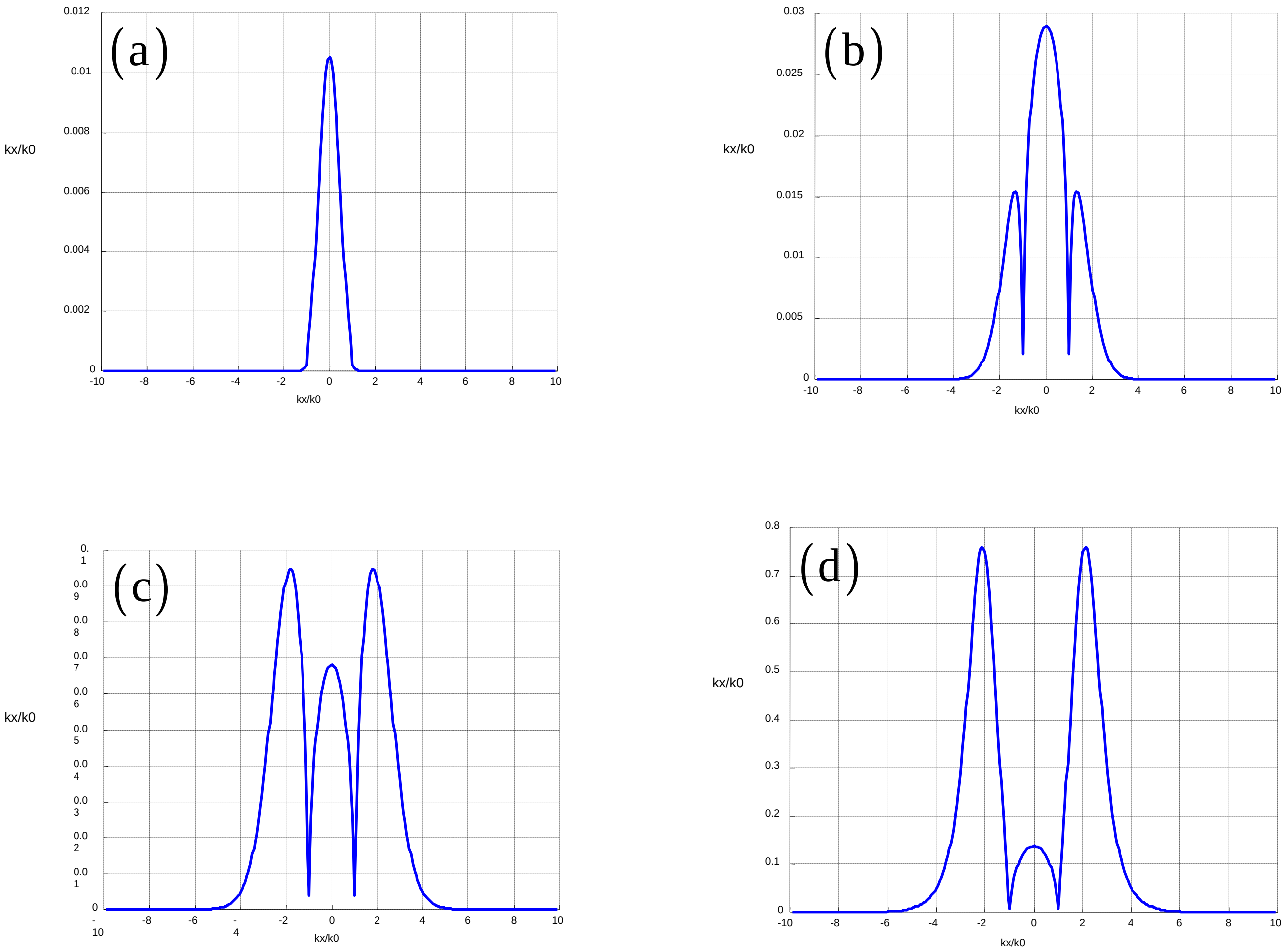
(a)
0.012
0.01
0.008
0.006
0.004
0.002
0
kx/k0
-10 -8 -6 -4 -2 0 2 4 6 8 10
kx/k0
(b)
0.03
0.025
0.02
0.015
0.01
0.005
0
kx/k0
-10 -8 -6 -4 -2 0 2 4 6 8 10
kx/k0
(c)
0.1
0.09
0.08
0.07
0.06
0.05
0.04
0.03
0.02
0.01
0
kx/k0
-10 -8 -6 -4 -2 0 2 4 6 8 10
kx/k0
(d)
0.8
0.7
0.6
0.5
0.4
0.3
0.2
0.1
0
kx/k0
-10 -8 -6 -4 -2 0 2 4 6 8 10
kx/k0

Fig.16

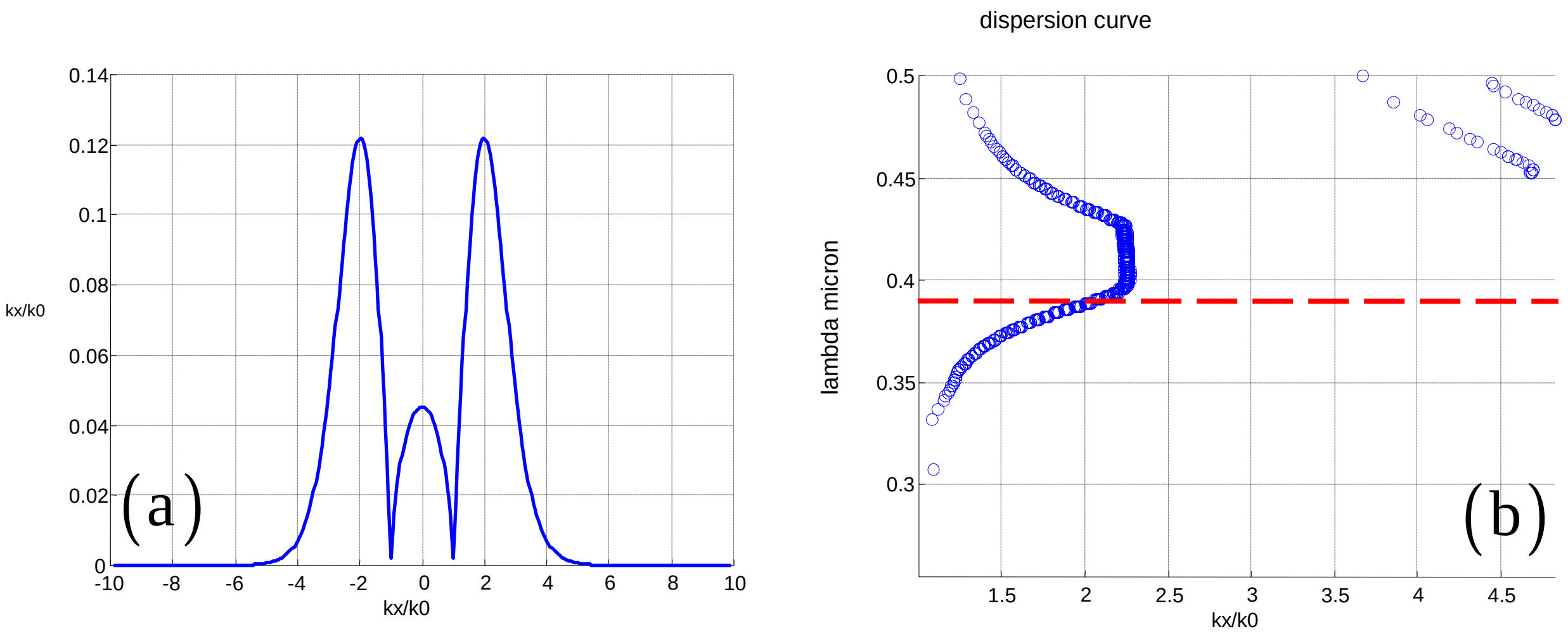

Fig.17

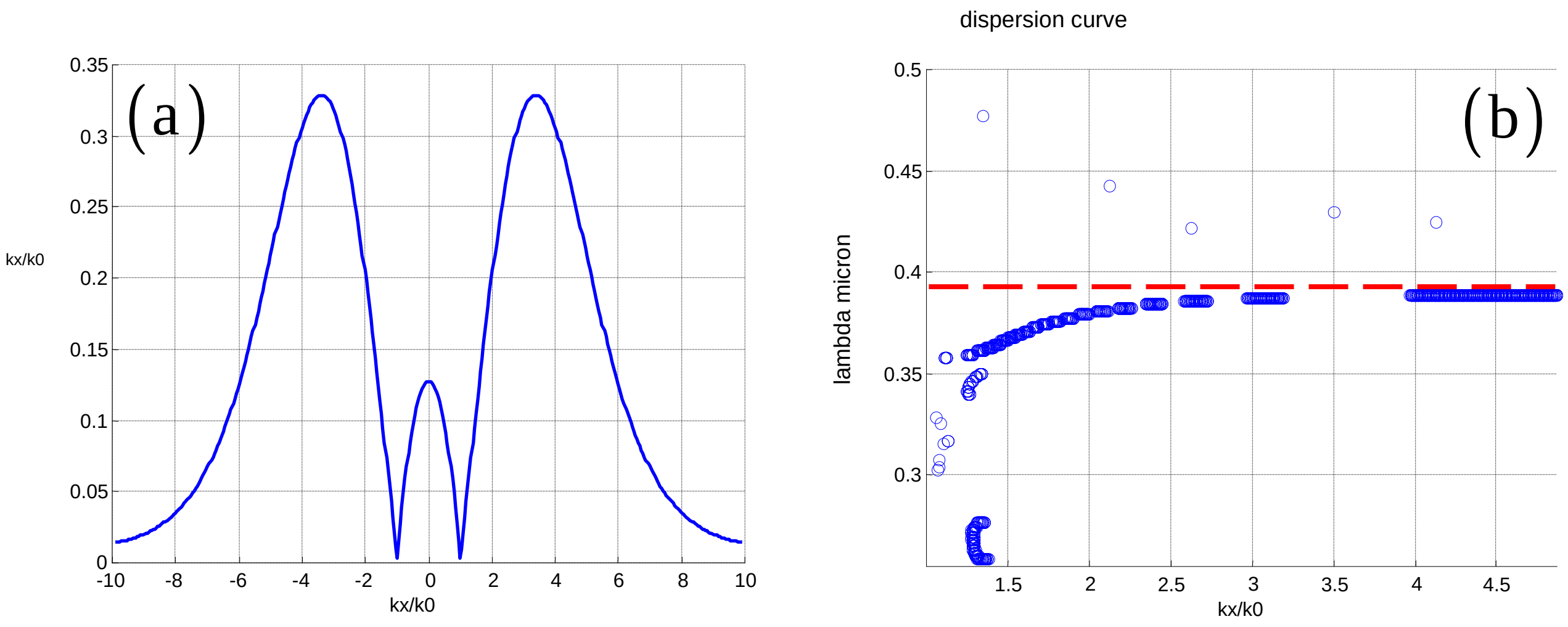

Pink: $n_{Ag}$=0.1824

Red: $n_{Ag}$=0.1500

Green: $n_{Ag}$=0.1000

Blue: $n_{Ag}$=0.0500

<u>$k_{Ag}$ is always 1.8164</u>

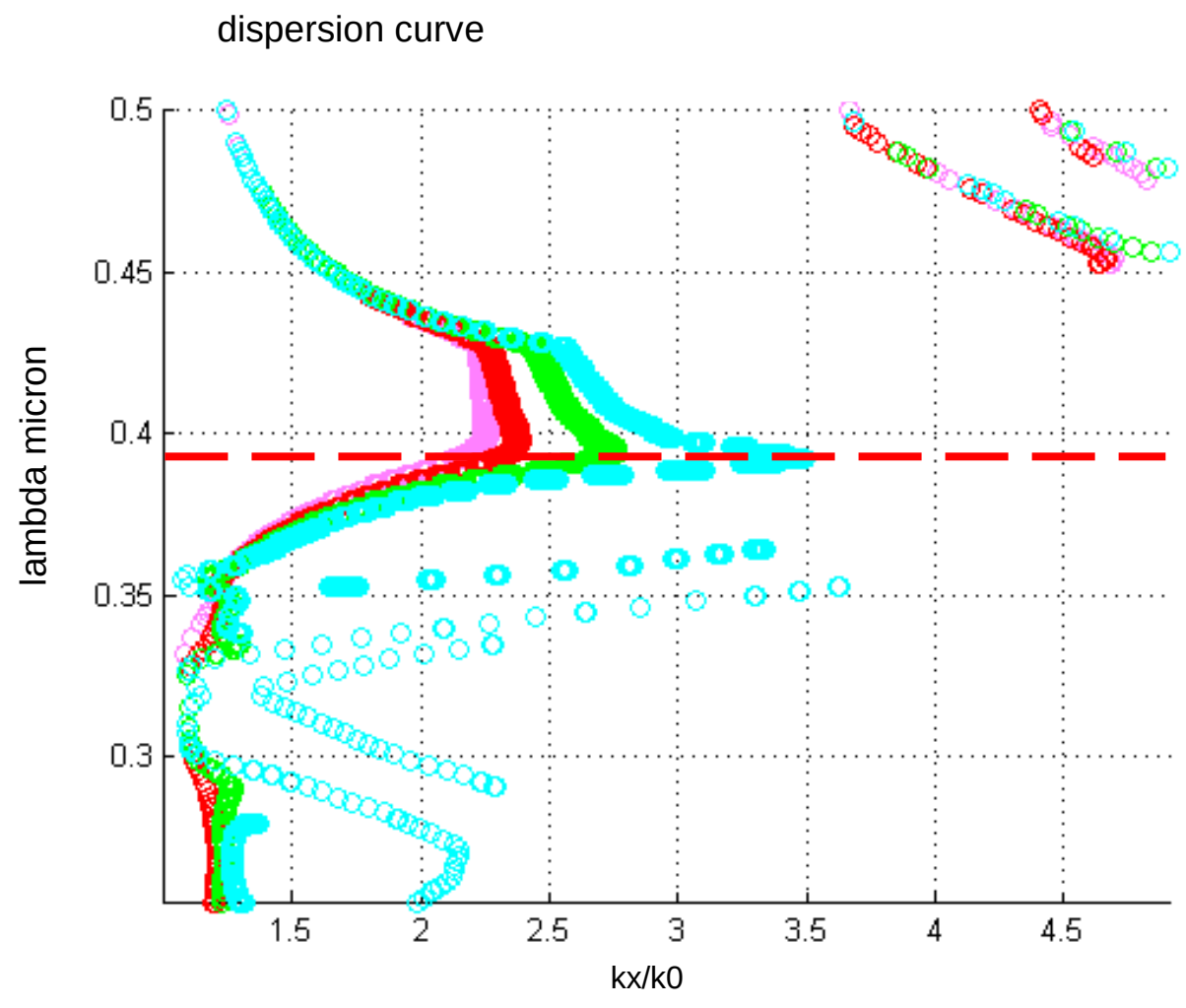

Fig.18

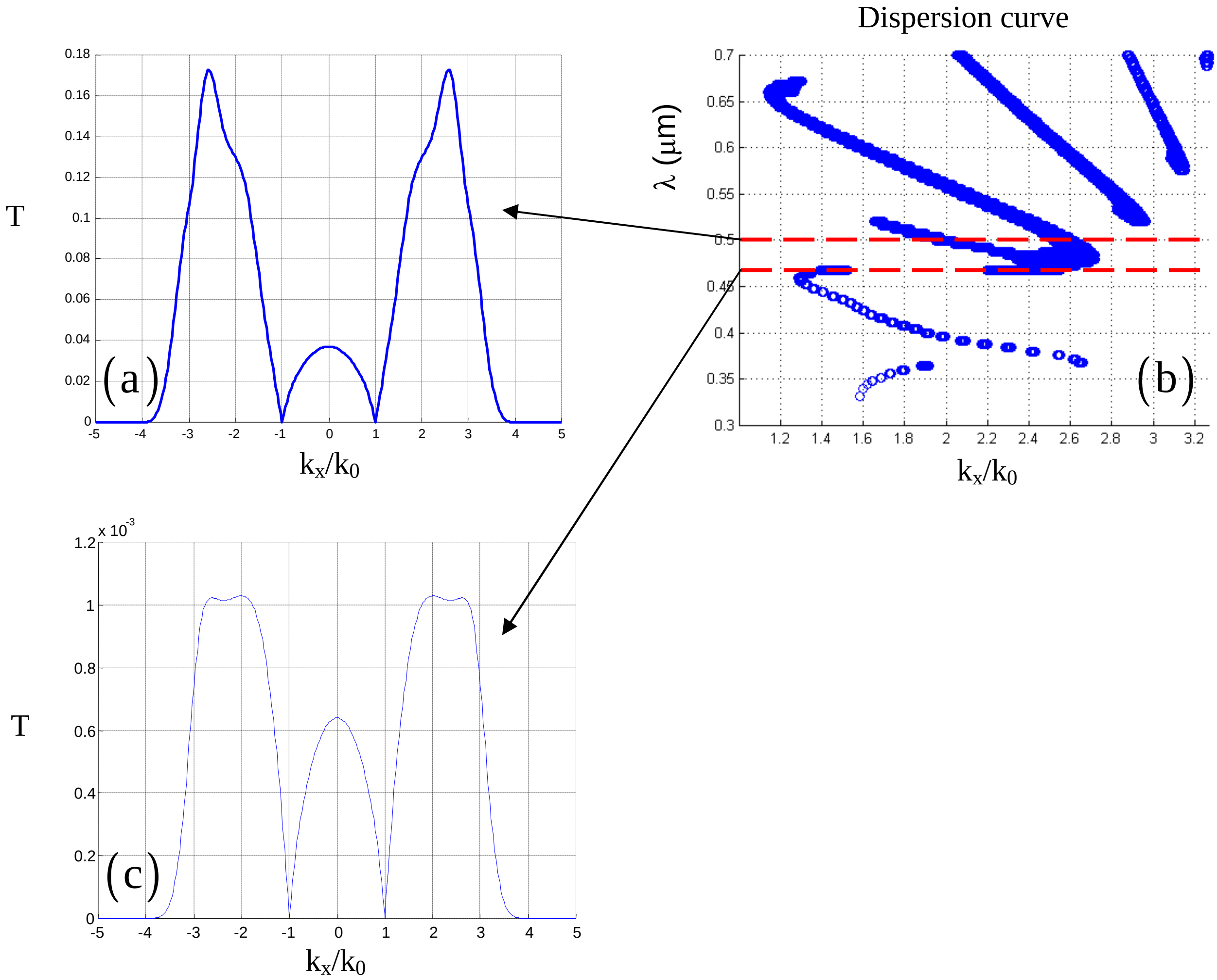

Fig.19

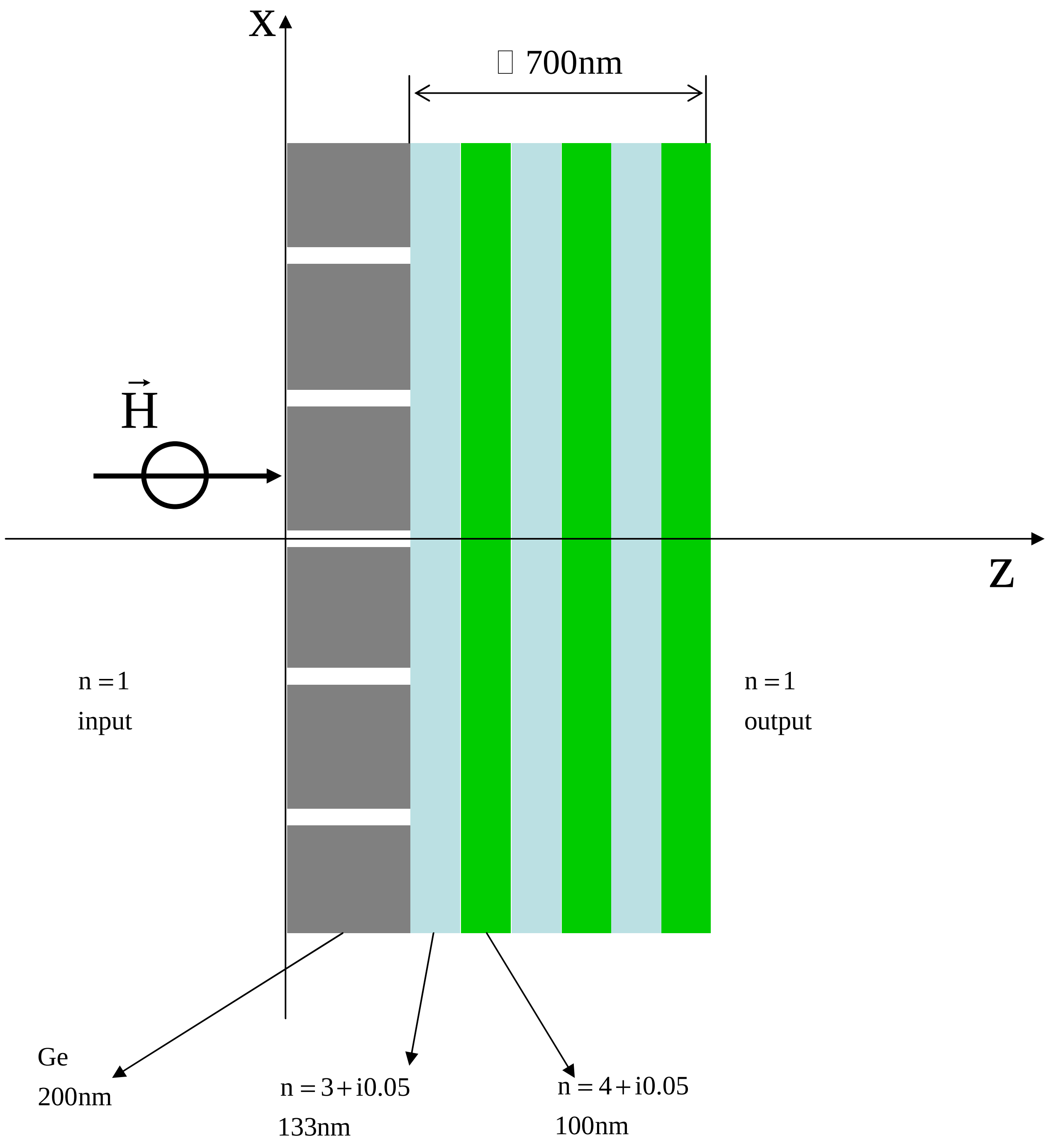

Fig.20

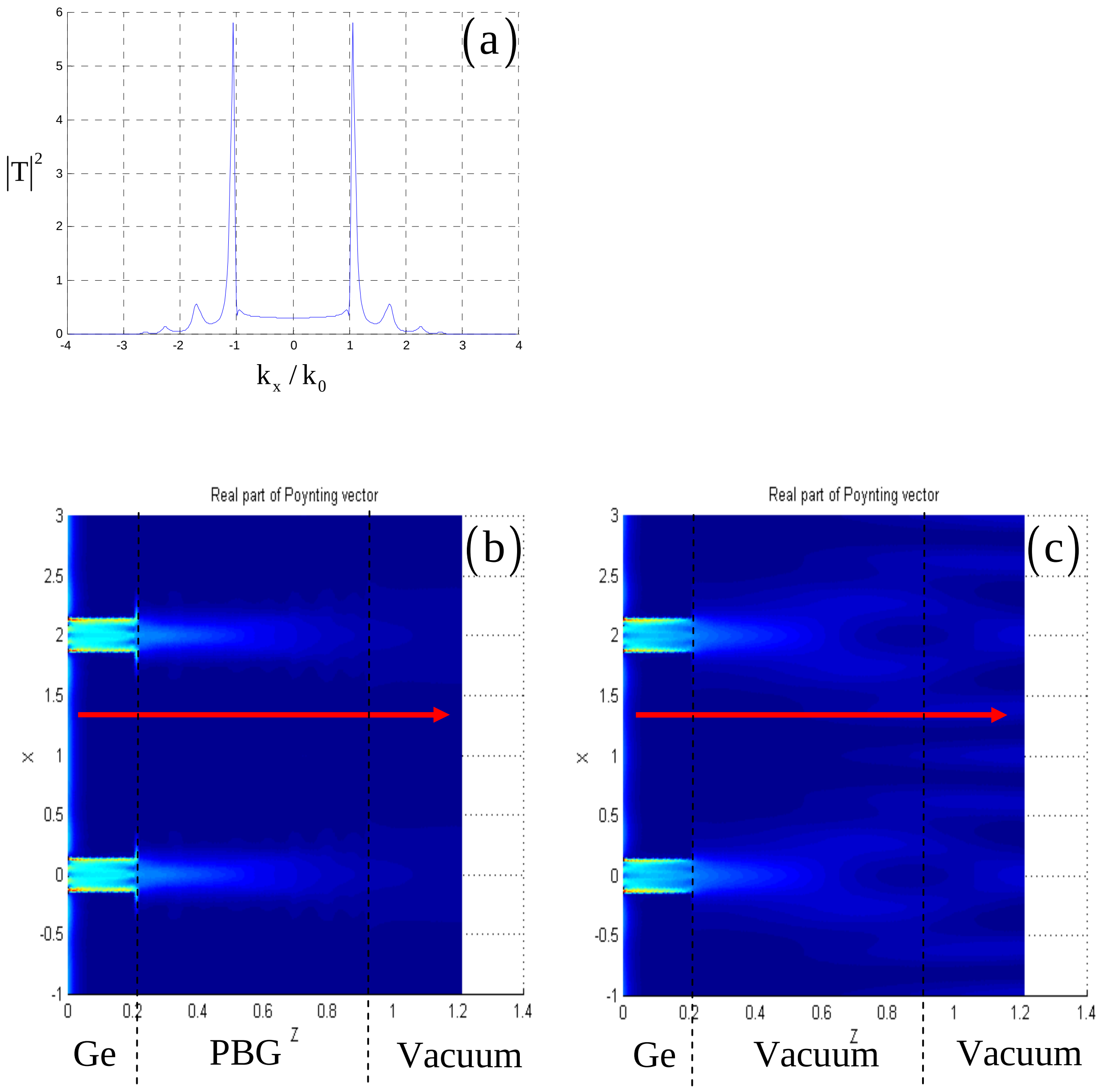

Fig.21